\documentclass[12pt,a4paper]{article}

\usepackage[english]{babel}
\usepackage[utf8]{inputenc}
\usepackage{amsmath,amsfonts}
\usepackage{graphicx}
\usepackage[margin=1in]{geometry}
\usepackage{enumitem}
\usepackage{comment}
\usepackage{setspace}
\usepackage{hyperref}
\usepackage[round]{natbib} 

\title{\textbf{Dynamics around the binary system (65803) Didymos}}

\date{
\small Proceedings IAU Symposium\\ 
Symposium Multi Scale (time and mass) Dynamics of Space Objects, Iaşi, Romania, 2021.\\
DOI: https://doi.org/10.1017/S1743921321001241\\
Published: 2021}

\begin{document}

\maketitle

\textbf{R. Machado Oliveira$^{1}$, O. C. Winter$^{1}$, R. Sfair$^{1}$, G. Valvano$^{1}$, T. S. Moura$^{1}$, G. Borderes-Motta$^{2}$.} \\

\footnotesize $^{1}$Grupo de Dinâmica Oribtal e Planetologia, São Paulo State University, UNESP, Guaratinguetá, CEP 12516-410, São Paulo, Brazil\\

\footnotesize $^{2}$Bioengineering and Aerospace Engineering Department, Universidad Carlos III de Madrid, Leganés, 28911, Madrid, Spain \\

\footnotesize First Author e-mail: rai.machado@unesp.br\\


\textbf{ABSTRACT}: Didymos and Dimorphos are primary and secondary, respectively, asteroids who compose a binary system that make up the set of Near Earth Asteroids (NEAs). They are targets of the Double Asteroid Redirection Test (DART), the first test mission dedicated to study of planetary defense, for which the main goal is to measure the changes caused after the secondary body is hit by a kinect impactor. The present work intends to  conduct a study, through numerical integrations, on the dynamics of massless particles distributed in the vicinity of the two bodies. An approximate shape for the primary body was considered as a model of mass concentrations (mascons) and the secondary was considered as a massive point.
Our results show the location and size of stable regions, and also their lifetime.\\

\vspace{0.2cm}

\setstretch{1.5}

\section{Introduction\label{S-intro}} 

The sub-kilometer asteroid Didymos and its moon Dimorphos form a binary system  
classified as a Near-Earth Asteroid and member of both the Apollo and Amor 
group\footnote{https://ssd.jpl.nasa.gov/sbdb.cgi?sstr=2065803}. 
The system was chosen as the target for the Double Asteroid Redirection Test (DART) mission, 
the first one dedicated to study of planetary defense \citep{cheng2018}. The main DART objective is to test 
the asteroid deflection technique by intentionally impacting Dimorphos and measuring the changes in its orbit. 
Ground-based observatories will also monitor these changes, and more precise 
information about the orbital evolution of Dimorphos after the impact will be assessed by the Hera 
mission\footnote{https://www.esa.int/Safety\_Security/Hera/Hera}.

Here we investigate the dynamics around the binary system, taking into account the 
irregular shape of the primary and also the gravitational disturbance of Dimorphos. 
Our goal is to analyze the orbital evolution of test particles 
in the vicinity of the system, and verify the existence of stable regions where particles can 
remain for an extended period and eventually pose a threat to the mission.

\cite{damme2017} made a similar study, but considering spacecraft (CubeSats) instead of natural particles. They adopted a gravitational model of Didymos, including spherical harmonics up to order and degree two, and integrating the system for short timescales (a few months). 
Here we consider a more complex gravitational model based on a polyhedron shape model and look for orbital stability on much longer timescales (several years). 

This paper is organized as follows: in Section~\ref{S-shape} we present the shape model adopted for Didymos, 
while the numerical setup and the main results are described in Section~\ref{S-stable}. 
The last section presents our final remarks.

\section{Shape model \label{S-shape}}

Given the irregular shape of Didymos, particles at the vicinity of the body are subject to a gravitational potential 
that cannot be reasonably modeled as originated from a point of mass or even an ellipsoid.  
However, the proper shape model for Didymos is not publicly available, and given that the object resembles the asteroid Ryugu, 
we assumed the shape as a scaled version of the latter. 

\begin{figure}[!h]
\centering
\includegraphics[width=6.5cm]{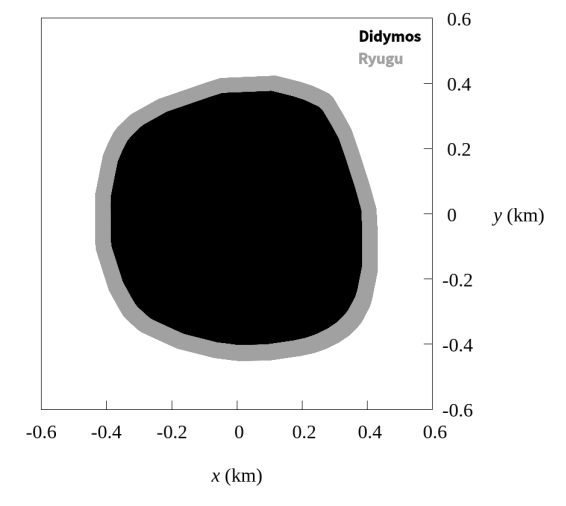}
\caption{Comparative size between Didymos and Ryugu model.}
\label{shape}
\end{figure}

From ground-based radar observations, \citet{scheirich2009} determined an equivalent diameter of 780~m 
and \citet{naidu2020} reported a bulk density as 2.17~gcm$^{-3}$ for the primary. 
Starting from a polyhedra representation of Ryugu composed of 574 vertices and 1144 triangular faces \citep{muller2011}, 
we determined the scaling factor that must be applied to Ryugu's shape to match the expected mass and volume for Didymos. 
A comparison between the two models is shown in Fig.~\ref{shape}.

Instead of computing the gravitational potential directly from the polyhedron \citep{werner1994}, 
our simulations were carried out with the more efficient, and yet precise, 
approach of mass concentrations -- mascons \citep{Geissler1996}, implemented in the N-BOM package \citep{Winter2020}. 
From a regularly spaced tridimensional grid that encompasses the object, the mascons are those grid points 
that lie within the shape model. Assuming that Didymos is homogeneous, 
the total mass of the asteroid is evenly distributed among the 26070 mascons. 

In our model, Didymos rotates with a period of 2.26~hours \citep{scheirich2009}. Dimorphos is represented as a point of 
mass with  3.45x$10^{9}$~kg orbiting the primary with a semimajor axis of 1178~m, and 
the eccentricity of 0.05 \citep{scheirich2009}. 

\section{Stable regions\label{S-stable}}

In order to explore the stability in the neighbourhood of the two bodies, we randomly distributed 20 thousand massless particles in a radial range between 0.45 and 1.45 km. 
The initial conditions of the particles were determined by keplerian orbits with eccentricity and inclination equal to zero.
The system was integrated for five years, which is the approximated interval between the DART\footnote{https://dart.jhuapl.edu/Mission/index.php} and HERA\footnote{https://www.heramission.space/} missions. 
The criterion for the remotion from the system was a collision or an ejection. For the collision we took into account the approximated dimension of the primary's mascons model and the estimated mean radius of the secondary given by \citet{scheirich2009}.
In the case of ejection, it was adopted 6 km as the maximum  radial distance from the primary, the same as considered in \citet{damme2017}. This distance is about five times the value of the semi-major axis of Dimorphos.

    \begin{figure}[!h]
    \centering
    \includegraphics[scale=.21]{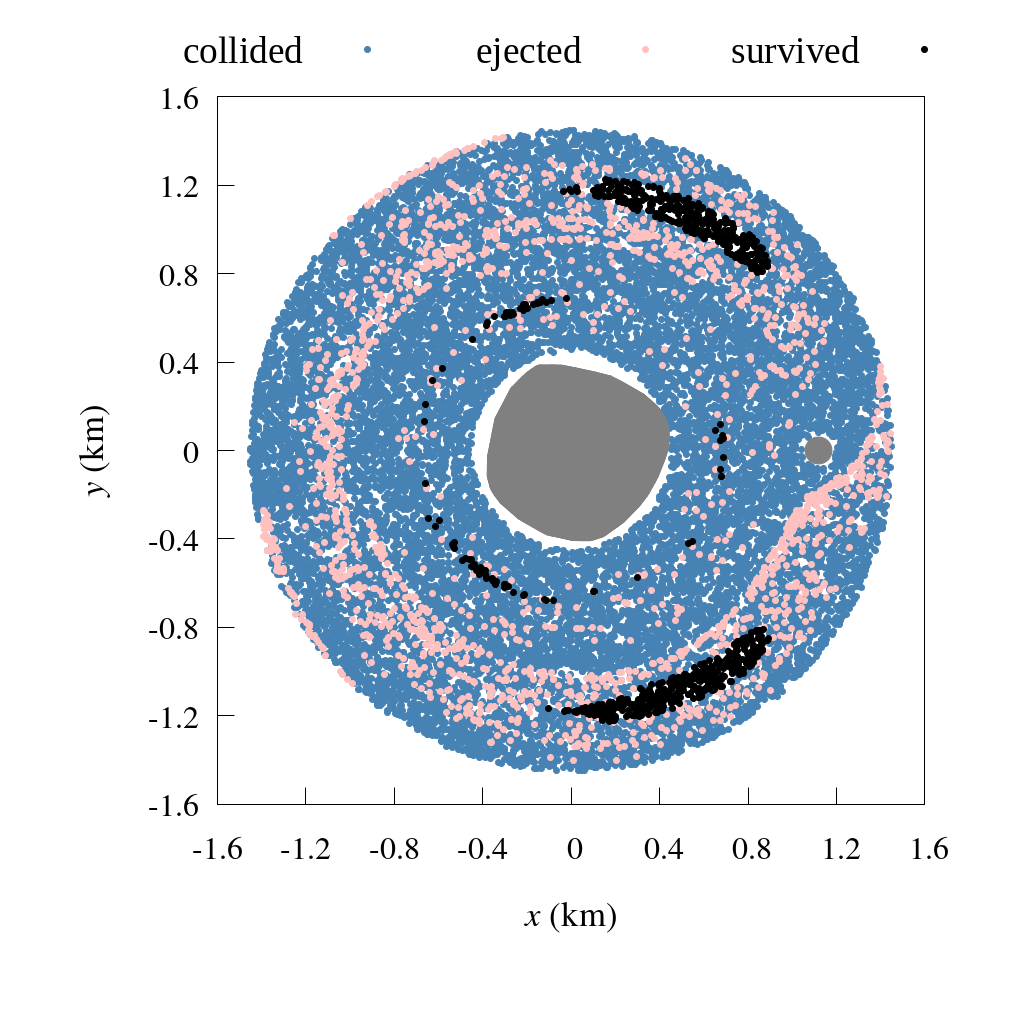}
    \includegraphics[scale=.21]{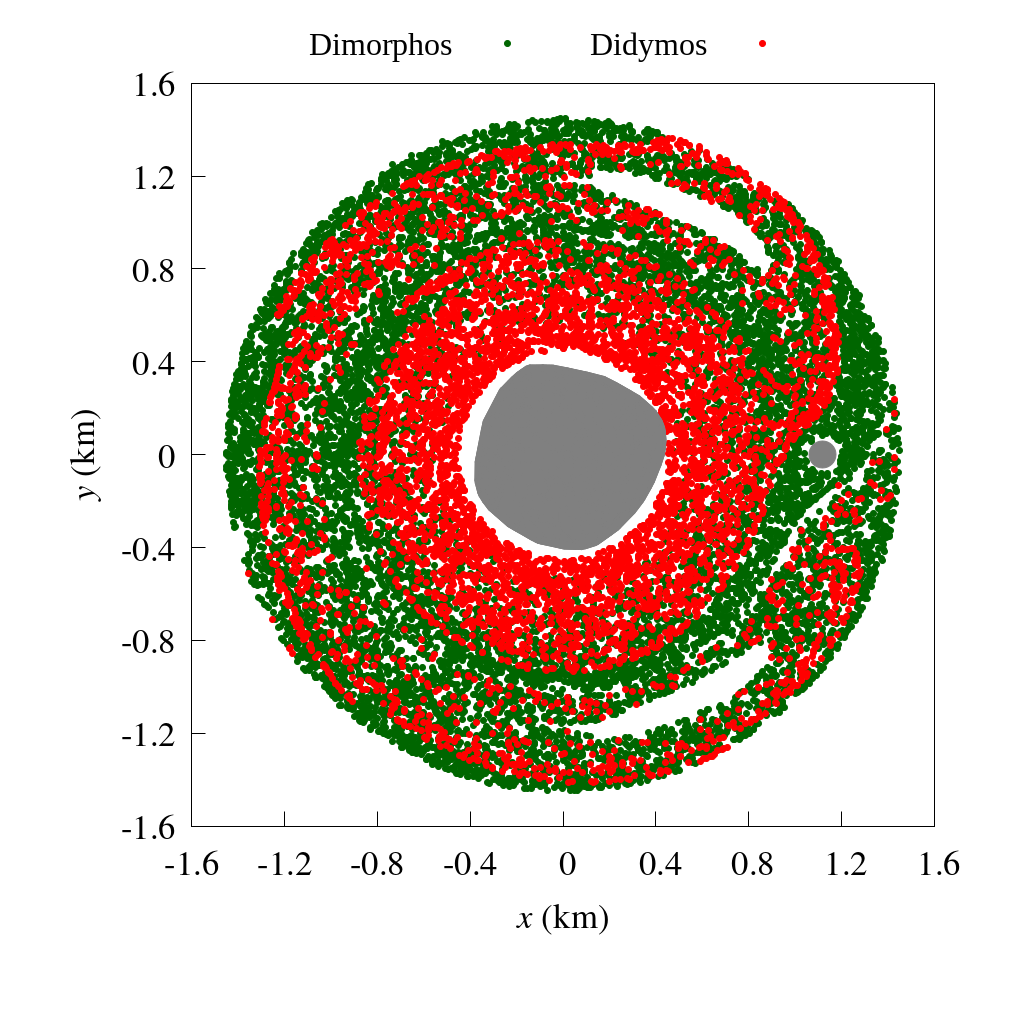}
    \caption{Fate of  the particle after 5 years of simulation. The particles are shown at their initial positions. Left: The blue dots represent the particles that collided with one of the bodies, the pink dots indicate  the ejected particles and the black dots the survived particles.
    Right: This plot shows only the collisions. The red dots represent the particles that collided with Didymos, while the green dots indicate the particles that collided with Dimorphos.}
    \label{sample-figure2}
    \end{figure} 
    
    The results are summarized in Fig. \ref{sample-figure2}. It shows the particles at their initial positions and the colors indicate the fate of the particles at the end of the simulations.
    After five years of simulations, 8.74\% were ejected, 30.3\% collided with Didymos, 56.94\% collided with Dimorphos and just 4.02\% survived.
    From the features of this plot can be made some considerations.
    The particles that collided are initially spread over several regions (left plot). 
    However, there is a pattern separating the the two groups of collision.
    As can be seen in the right plot, the particles that collided with Dimorphos are mainly located in  two spiral arms that depart from the satellite. In the case of particles that collided with Didymos, they can be separated into two groups: one in a disk inside the orbit of Dimorphos and other involving the stable regions around the triangular lagrangian points. 
    On the other hand, the particles that were ejected and those that survived are found initially in specific regions. The ejected particles were preferentially in places close to the orbit of Dimorphos. Their fate was determined by close encounters with the secondary. 
    
\begin{figure}[!h]
    \centering
    \includegraphics[scale=.21]{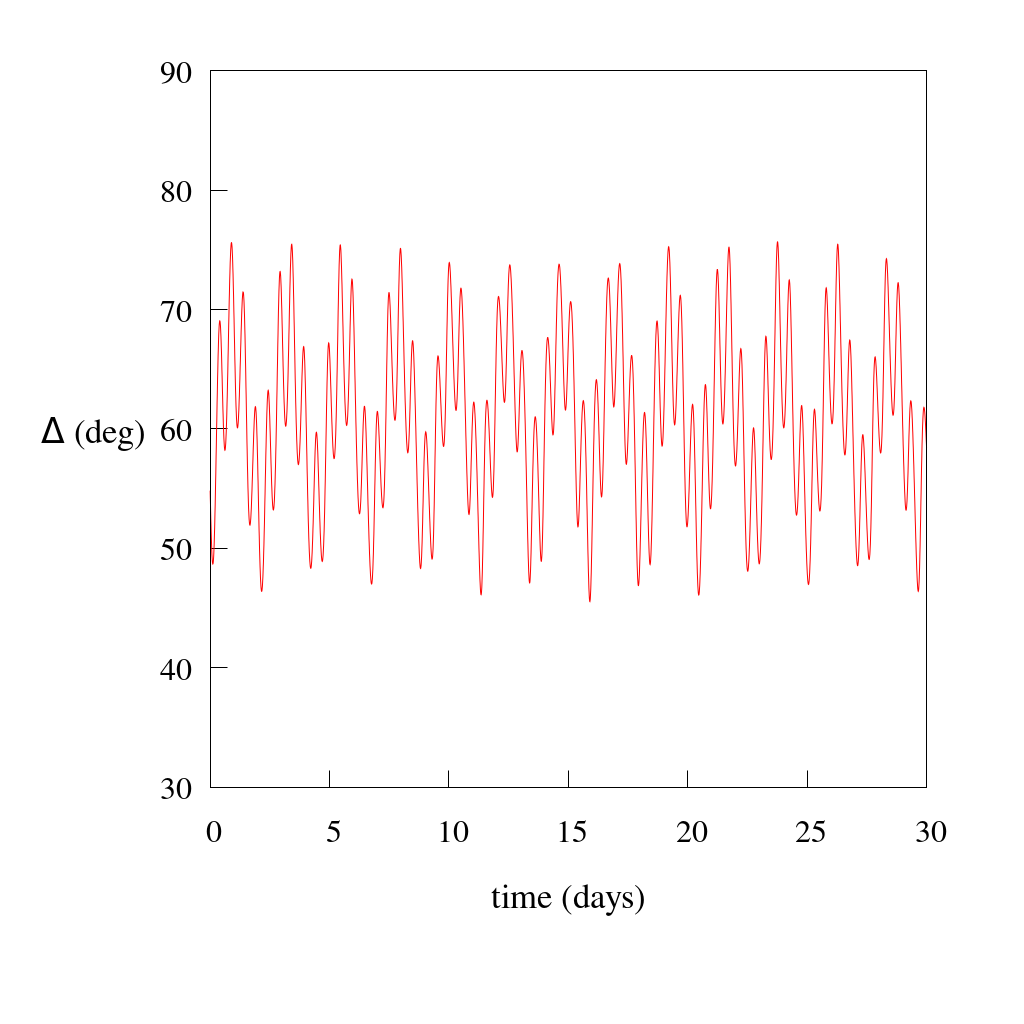}
    \includegraphics[scale=.21]{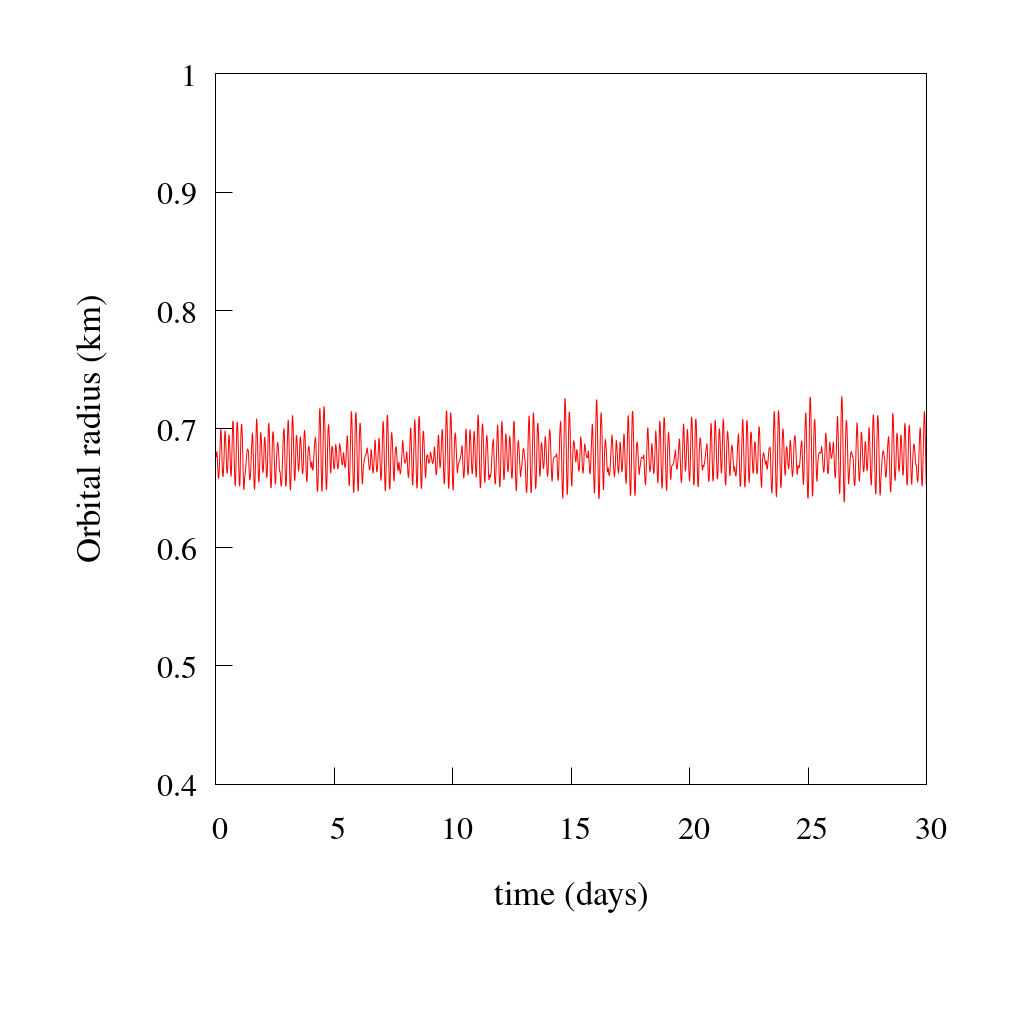}
    \caption{Left: Example of the evolution of the angle between Dimorphos and a particle located around $L_4$ lagrangian point. Right: Example of the radial evolution of the trajectory of a particle located in one of the stable arcs close to Didymos.  }
    \label{sample-figure4}
    \end{figure}

    In the case of the survivors, Fig. \ref{sample-figure2} (left plot) shows that they are initially located in the coorbital region, around the lagrangian points  $L_4$ and $L_5$, and also in arcs of a ring formed between the two bodies.
    The particles around one of the two triangular lagrangian points present behaviours expected in binary systems.  Since these are stable equilibrium points for the given mass ratio of the binary, 
    it is coherent to have survivors around them. These particles show a tadpole like trajectory, as in the example given in  Fig. \ref{sample-figure4} (left plot).
    The trajectories of the particles in the stable regions close to Didymos show a very narrow amplitude of radial variation. An example is given in Fig. \ref{sample-figure4} (right plot).   
    
\begin{figure}[!h]
    \centering
    \includegraphics[scale=.23]{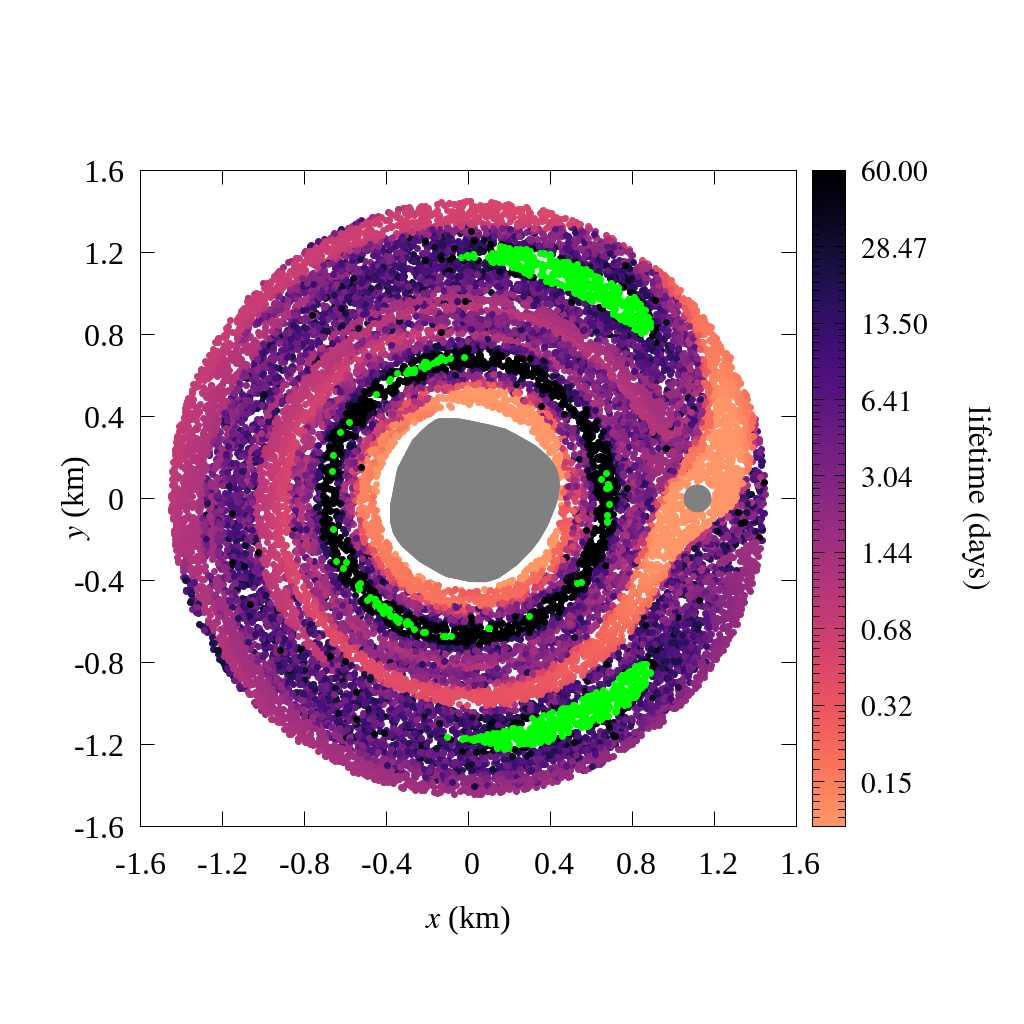}
    \caption{Lifetime of the particles according to their initial location. 
    The particles that survived the whole integration time (5 years) are indicated in green.}
    \label{sample-figure3}
    \end{figure}

    An idea of the time evolution of the system as a whole can be seen in Fig. \ref{sample-figure3}.
    This plot shows the lifetime of the particles according to their initial location. The survivors are indicated in green. 
    The system as seen in Fig. \ref{sample-figure2} is quickly defined, since 79\% of the particles are removed within 10 days. 
    Those that were removed in just a few hours are the ones that were close to the surface of Didymos and those that were in spiral arms associated to Dimorphos. The satellite is the responsible for the large majority of this fast removal in the first orbital periods of the particles, by collision with Dimorphos or ejection of the region.   
    
    Note in the Fig. \ref{sample-figure3} that the particles which live longer (very dark color) are in the neighbourhood of the survivors (in green). Some of these particles can stay alive for even more than one year of integration.
    The orbital period of the particles in the dark ring shown in Fig. \ref{sample-figure3} is a value in the range between five and six hours.
    
    A comparison of these results with those of  \cite{damme2017} shows that the general structure is similar. However, in their work was found a large stable ring of initial conditions close to Didymos which is not the case in our results. That is certainly due to the large difference in timescales considered (a few months for them, while  some years in ours), but the effects due to different gravitational potential models also contributed in the evolution of the trajectories with  initial conditions closer to Didymos.

\section{Final comments \label{S-comments}}

In this work we explored the long term stability of massless particles in the Didymos-Dimorphos system.
In order to take into account the gravitational contribution of Didymos, we developed an approximated shape model in terms of a polyhedron  composed of triangular faces and then, transformed it in a model of Mascons.

The results clearly show the location and size of the stable regions. 
An analysis of the lifetime of the particles indicated that the huge majority of the particles with unstable trajectories are removed in just ten days.
The largest amount of particles in the stable regions have tadpole like trajectories around the triangular lagrangian equilibrium points.

\section{Acknowledgements}

This study was financed in part by the Coordenação de Aperfeiçoamento de Pessoal de Nível Superior - Brasil (CAPES) - Finance Code 001, Fundação de Amparo à Pesquisa do Estado de São Paulo (FAPESP) - Proc. 2016/24561-0 and Proc. 2020/14307-4, Conselho Nacional de Desenvolvimento Cientifico e Tecnológico (CNPq) - Proc. 120338/2020-3 and Proc. 305210/2018-1

\bibliographystyle{apalike}
\bibliography{references}

\end{document}